\documentclass[longauth]{aa} % for the long lists of affiliations 
\usepackage{graphicx}
\usepackage{txfonts}
\usepackage{makecell}
\usepackage{placeins}

\usepackage[usenames, dvipsnames]{color}
\newcommand{\Adam}{}

\begin{document}

   \title{Glancing through the debris disk: Photometric analysis of DE Boo with CHEOPS\thanks{This article uses data from CHEOPS programme CH\_PR100010.}}

   \author{Á. Boldog\inst{1,2,3}, 
Gy. M. Szabó\inst{4,3}, 
L. Kriskovics\inst{1,2}, 
A. Brandeker\inst{5}, 
F. Kiefer\inst{6,7}, 
A. Bekkelien\inst{8}, 
P. Guterman\inst{9,10}, 
G. Olofsson\inst{5}, 
A. E. Simon\inst{11}, 
D. Gandolfi\inst{12}, 
L. M. Serrano\inst{12}, 
T. G. Wilson\inst{13}, 
S. G. Sousa\inst{14}, 
A. Lecavelier des Etangs\inst{15}, 
Y. Alibert\inst{11}, 
R. Alonso\inst{16,17}, 
G. Anglada\inst{18,19}, 
T. Bandy\inst{20}, 
T. Bárczy\inst{21}, 
D. Barrado\inst{22}, 
S. C. C. Barros\inst{14,23}, 
W. Baumjohann\inst{24}, 
M. Beck\inst{8}, 
T. Beck\inst{11}, 
W. Benz\inst{11,25}, 
N. Billot\inst{8}, 
X. Bonfils\inst{26}, 
C. Broeg\inst{11,27}, 
M. Buder\inst{28}, 
J. Cabrera\inst{29}, 
S. Charnoz\inst{30}, 
A. Collier Cameron\inst{13}, 
C. Corral van Damme\inst{31}, 
Sz. Csizmadia\inst{29}, 
M. B. Davies\inst{32}, 
A. Deline\inst{8}
M. Deleuil\inst{9}, 
L. Delrez\inst{33,34}, 
O. D. S. Demangeon\inst{14,23}, 
B.-O. Demory\inst{25}, 
D. Ehrenreich\inst{8}, 
A. Erikson\inst{29}, 
J. Farinato\inst{35}, 
A. Fortier\inst{11,25}, 
L. Fossati\inst{24}, 
M. Fridlund\inst{36,37}, 
M. Gillon\inst{33}, 
M. Güdel\inst{38}, 
K. Heng\inst{25,39}, 
S. Hoyer\inst{9}, 
K. G. Isaak\inst{40}, 
L. L. Kiss\inst{1,2,41}, 
J. Laskar\inst{42}, 
M. Lendl\inst{8}, 
C. Lovis\inst{8}, 
D. Magrin\inst{43}, 
P. F. L. Maxted\inst{44}, 
M. Mecina\inst{45}, 
V. Nascimbeni\inst{43}, 
R. Ottensamer\inst{45}, 
I. Pagano\inst{35}, 
E. Pallé\inst{16}, 
G. Peter\inst{28}, 
G. Piotto\inst{43,46}, 
D. Pollacco\inst{39}, 
D. Queloz\inst{47,48}, 
R. Ragazzoni\inst{43,46}, 
N. Rando\inst{31}, 
H. Rauer\inst{29,49,50}, 
I. Ribas\inst{18,19}, 
N. C. Santos\inst{14,23}, 
G. Scandariato\inst{35}, 
D. Ségransan\inst{8}, 
A. M. S. Smith\inst{29}, 
M. Steller\inst{24}, 
N. Thomas\inst{11}, 
S. Udry\inst{8}, 
V. Van Grootel\inst{34}, 
N. A. Walton\inst{51}}

   \institute{$^{1}$\label{inst:1} Konkoly Observatory, Research Centre for Astronomy and Earth Sciences, 1121 Budapest, Konkoly Thege Miklós út 15-17, Hungary \and
\label{inst:2} CSFK, MTA Centre of Excellence, Budapest, Konkoly Thege Miklós út 15-17., H-1121, Hungary \and
\label{inst:3} MTA-ELTE Exoplanet Research Group, 9700 Szombathely, Szent Imre h. u. 112, Hungary \and
\label{inst:4} ELTE E\"otv\"os Lor\'and University, Gothard Astrophysical Observatory, 9700 Szombathely, Szent Imre h. u. 112, Hungary \and
\label{inst:5} Department of Astronomy, Stockholm University, AlbaNova University Center, 10691 Stockholm, Sweden \and
\label{inst:6} Sorbonne Université, CNRS, UMR 7095, Institut d’Astrophysique de Paris, 98 bis bd Arago, 75014 Paris, France \and
\label{inst:7} LESIA, Observatoire de Paris, Université PSL, CNRS, Sorbonne Université, Université Paris Cité, 5 place Jules Janssen, 92195Meudon, France \and
\label{inst:8} Observatoire Astronomique de l'Université de Genève, Chemin Pegasi 51, CH-1290 Versoix, Switzerland \and
\label{inst:9} Aix Marseille Univ, CNRS, CNES, LAM, 38 rue Frédéric Joliot-Curie, 13388 Marseille, France \and
\label{inst:10} Division Technique INSU, CS20330, 83507 La Seyne sur Mer cedex, France \and
\label{inst:11} Physikalisches Institut, University of Bern, Sidlerstrasse 5, 3012 Bern, Switzerland \and
\label{inst:12} Dipartimento di Fisica, Universita degli Studi di Torino, via Pietro Giuria 1, I-10125, Torino, Italy \and
\label{inst:13} Centre for Exoplanet Science, SUPA School of Physics and Astronomy, University of St Andrews, North Haugh, St Andrews KY16 9SS, UK \and
\label{inst:14} Instituto de Astrofisica e Ciencias do Espaco, Universidade do Porto, CAUP, Rua das Estrelas, 4150-762 Porto, Portugal \and
\label{inst:15} Institut d'astrophysique de Paris, UMR7095 CNRS, Université Pierre \& Marie Curie, 98bis blvd. Arago, 75014 Paris, France \and
\label{inst:16} Instituto de Astrofisica de Canarias, 38200 La Laguna, Tenerife, Spain \and
\label{inst:17} Departamento de Astrofisica, Universidad de La Laguna, 38206 La Laguna, Tenerife, Spain \and
\label{inst:18} Institut de Ciencies de l'Espai (ICE, CSIC), Campus UAB, Can Magrans s/n, 08193 Bellaterra, Spain \and
\label{inst:19} Institut d'Estudis Espacials de Catalunya (IEEC), 08034 Barcelona, Spain \and
\label{inst:20} Physikalisches Institut, University of Bern, Switzerland \and
\label{inst:21} Admatis, 5. Kandó Kálmán Street, 3534 Miskolc, Hungary \and
\label{inst:22} Depto. de Astrofisica, Centro de Astrobiologia (CSIC-INTA), ESAC campus, 28692 Villanueva de la Cañada (Madrid), Spain \and
\label{inst:23} Departamento de Fisica e Astronomia, Faculdade de Ciencias, Universidade do Porto, Rua do Campo Alegre, 4169-007 Porto, Portugal \and
\label{inst:24} Space Research Institute, Austrian Academy of Sciences, Schmiedlstrasse 6, A-8042 Graz, Austria \and
\label{inst:25} Center for Space and Habitability, University of Bern, Gesellschaftsstrasse 6, 3012 Bern, Switzerland \and
\label{inst:26} Université Grenoble Alpes, CNRS, IPAG, 38000 Grenoble, France \and
\label{inst:27} Center for Space and Habitability, Gesellsschaftstrasse 6, 3012 Bern, Switzerland \and
\label{inst:28} Institute of Optical Sensor Systems, German Aerospace Center (DLR), Rutherfordstrasse 2, 12489 Berlin, Germany \and
\label{inst:29} Institute of Planetary Research, German Aerospace Center (DLR), Rutherfordstrasse 2, 12489 Berlin, Germany \and
\label{inst:30} Université de Paris, Institut de physique du globe de Paris, CNRS, F-75005 Paris, France \and
\label{inst:31} ESTEC, European Space Agency, 2201AZ, Noordwijk, NL \and
\label{inst:32} Centre for Mathematical Sciences, Lund University, Box 118, 221 00 Lund, Sweden \and
\label{inst:33} Astrobiology Research Unit, Université de Liège, Allée du 6 Août 19C, B-4000 Liège, Belgium \and
\label{inst:34} Space sciences, Technologies and Astrophysics Research (STAR) Institute, Université de Liège, Allée du 6 Août 19C, 4000 Liège, Belgium \and
\label{inst:35} INAF, Osservatorio Astrofisico di Catania, Via S. Sofia 78, 95123 Catania, Italy \and
\label{inst:36} Leiden Observatory, University of Leiden, PO Box 9513, 2300 RA Leiden, The Netherlands \and
\label{inst:37} Department of Space, Earth and Environment, Chalmers University of Technology, Onsala Space Observatory, 439 92 Onsala, Sweden \and
\label{inst:38} University of Vienna, Department of Astrophysics, Türkenschanzstrasse 17, 1180 Vienna, Austria \and
\label{inst:39} Department of Physics, University of Warwick, Gibbet Hill Road, Coventry CV4 7AL, United Kingdom \and
\label{inst:40} Science and Operations Department - Science Division (SCI-SC), Directorate of Science, European Space Agency (ESA), European Space Research and Technology Centre (ESTEC), Keplerlaan 1, 2201-AZ Noordwijk, The Netherlands \and
\label{inst:41} ELTE E\"otv\"os Lor\'and University, Institute of Physics, P\'azm\'any P\'eter s\'et\'any 1/A, 1117 \and
\label{inst:42} IMCCE, UMR8028 CNRS, Observatoire de Paris, PSL Univ., Sorbonne Univ., 77 av. Denfert-Rochereau, 75014 Paris, France \and
\label{inst:43} INAF, Osservatorio Astronomico di Padova, Vicolo dell'Osservatorio 5, 35122 Padova, Italy \and
\label{inst:44} Astrophysics Group, Keele University, Staffordshire, ST5 5BG, United Kingdom \and
\label{inst:45} Department of Astrophysics, University of Vienna, Tuerkenschanzstrasse 17, 1180 Vienna, Austria \and
\label{inst:46} Dipartimento di Fisica e Astronomia "Galileo Galilei", Universita degli Studi di Padova, Vicolo dell'Osservatorio 3, 35122 Padova, Italy \and
\label{inst:47} ETH Zurich, Department of Physics, Wolfgang-Pauli-Strasse 2, CH-8093 Zurich, Switzerland \and
\label{inst:48} Cavendish Laboratory, JJ Thomson Avenue, Cambridge CB3 0HE, UK \and
\label{inst:49} Zentrum für Astronomie und Astrophysik, Technische Universität Berlin, Hardenbergstr. 36, D-10623 Berlin, Germany \and
\label{inst:50} Institut für Geologische Wissenschaften, Freie Universität Berlin, 12249 Berlin, Germany \and
\label{inst:51} Institute of Astronomy, University of Cambridge, Madingley Road, Cambridge, CB3 0HA, United Kingdom 
}

   \date{Received ; accepted }

% \abstract{}{}{}{}{} 
% 5 {} token are mandatory
 
  \abstract
   {}
   {DE Boo is a unique system, with an edge-on view through the debris disk around the star. The disk, which is analogous to the Kuiper belt in the Solar System, was reported to extend from 74 to 84 AU from the central star. The high photometric precision of the Characterising Exoplanet Satellite (CHEOPS) provided an exceptional opportunity to observe small variations in the light curve due to transiting material in the disk. This is a unique chance to investigate processes in the debris disk.}
   {Photometric observations of DE Boo of a total of four days were carried out with CHEOPS. Photometric variations due to spots on the stellar surface were subtracted from the light curves by applying a two-spot model and a fourth-order polynomial. The photometric observations were accompanied by spectroscopic measurements with the 1m RCC telescope at Piszkéstető and with the SOPHIE spectrograph in order to refine the astrophysical parameters of DE Boo.}
   {We present a detailed analysis of the photometric observation of DE Boo. We report the presence of nonperiodic transient features in the residual light curves with a transit duration of 0.3--0.8 days. We calculated the maximum distance of the material responsible for these variations to be 2.47 AU from the central star, much closer than most of the mass of the debris disk. Furthermore, we report the first observation of flaring events in this system.}
   {We interpreted the transient features as the result of scattering in an inner debris disk around DE Boo. The processes responsible for these variations were investigated in the context of interactions between planetesimals in the system.}

    \authorrunning{Boldog et al.}
   \maketitle
%
%-------------------------------------------------------------------

\section{Introduction}

Debris disks have been observed outside of our Solar System since the 1980s \citep{Aumann1984, Smith1984}. In contrast to primordial disks of gas and dust, which dissipate after a few million years of the formation of the star, dusty debris disks can be observed for a longer period even around K type stars \citep{2006ApJ...644..525M}. These disks are made of dust that experiences frequent collisions with larger bodies such as comets or planetesimals along their orbits. The semi major axes of debris disks differ from system to system. Some debris disks are found as close to the central star as around 1~AU, such as in the HD 69830 system \citep{Lisse2007}, while others can stretch out several hundreds of AU \citep{Smith1984, Kalas2007}. Systems with both an inner and an outer debris disk have also been reported \citep{Lisse2008, Stark2009}.

High-sensitivity photometry can provide unique information on the structure of an edge-on debris disk. When disk inhomogeneities pass in front of the star, they produce variations in the star's extinction, dimming the stellar light. These variations are interpreted as transits of  structures in the circumstellar disk  \citep{1999A&A...343..916L}. 

During its nominal mission, TESS also observed quasi-periodic and quasi-stochastic transients in ``dipper’’ stars, which are surrounded by debris disks \citep{Gaidos2019}. The observed features show a quasi-continuous spectrum of separated, nonrecurrent random transits to quasi-periodic events with a period of 1--4 days. In the case of HD 240779 for example, the dimming vanished over 2 years between two TESS visits, suggesting that the origin of the transient was disrupted or underwent efficient evaporation \citep{Gaidos2022}. {\Adam Most of these dippers are early-type stars,} while K-M stars can also be found between them.
Another famous example for an edge-on disk is the $\beta$ Pic system \citep{Smith1984}, where transient dips have been observed \citep{2019A&A...625L..13Z,Pavlenko2022,2022NatSR..12.5855L}, with a flux drop at the level of 0.03\%{} at maximum and a duration of less than 1 day, and with no periodicity seen in the sectors investigated. Unlike the dipper stars, the transients in the $\beta$ Pic system were attributed to exocomets.

Ground-based facilities are limited in their ability to observe these variations because of the Earth's atmosphere and the diurnal observation gaps. We started a space-based photometry program with the Characterising Exoplanet Satellite (CHEOPS) to detect these tiny variations in
a few selected systems. The CHEOPS space telescope launched in 2019 and designed to detect and characterise the transits of small exoplanets \citep{2021ExA....51..109B}, is well placed to perform this search because of its high photometric precision \citep{Maxted2021}.

DE Bootis (HD 131511) is a chromospherically active spectroscopic binary (SB 1)  dwarf RS\,CVn system as identified by \citet{Eker2008}, with a K0V primary component \citep{Gray2003}. The stellar fundamental parameters for the primary and the M dwarf companion are shown in Table \ref{tab:star_param}. The binary has a separation of 0.19$\pm$0.03 AU, an orbital period of 125.396 days with the eccentricity of the orbit being $e = 0.51\pm0.01$ \citep{Kennedy2014}. The age of the system was calculated by \citet{Gray2015} to be less {\Adam than} 797 Myr, while \cite{Marshall2014} found it to be between 570 and 700 Myr. DE Boo exhibits signs of magnetic activity, such as emission in the Ca II H and K lines and large stellar spots. These spots can be observed in the rotational modulation of the light curve with a period of 10.39$^\mathrm{d}$ \citep{Henry1995}. A radially narrow debris disk at around 70 AU from the primary component has been reported by \cite{Marshall2014}. DE Boo is among the closest systems with
a debris disk, lying at 37.928$\pm$0.243 light years from Earth \citep{GAIAEDR3}, while its Gaia DR2 magnitude is 5.766, leading to a 19.92 ppm internal precision per hour with CHEOPS.
The disk, analogous to the Solar System's Kuiper-belt, is aligned with the orbit of the host binary and has a near edge-on geometry, as derived by \citet{Jancart2005} using Hipparcos data. Furthermore, \citet{Marshall2014} showed that the inclination of the disk is at least 84$^\circ$. Due to this, the debris disk is always in the line of sight, continuously causing dips in the light curve.

The level of the expected variations can be estimated via the value of $f=L_\mathrm{disk}/L_\mathrm{star}$, which implies that $f=4\cdot 10^{-6}$ of the starlight is intercepted by the dust. The radial optical depth is $\tau = f/\sin(\theta/2)$, where theta is the opening angle
\citep{2014MNRAS.438.3299K}. The resulting $\tau\approx1\cdot 10^{-5}$ is roughly the level of variation expected
if there are radial holes in the dust distribution. 
Because of the frequent collisions between the smallest bodies inside the disk and the Keplerian shear, such holes are unexpected. However, a similar process was suggested in the case of dippers \citep{2020ApJS..251...18T}. The transit of planetesimals that are collecting dense dust clumps inside their Hill radius can still result in detectable light variations.
The viewing geometry of DE Boo along with the high expected S/N make this system an ideal target for the Dusty Debris Disk (DDD) survey with CHEOPS, and we have chosen this system as one of the targets of our program.

The co-planarity of the debris disk with the orbit of the binary was studied by \citet{Kennedy2014}.  The favorable orientation allows us to examine the behavior of the material in the debris disk ---such as extrasolar collisions--- early in the evolution of the system. The outstanding precision of CHEOPS makes DE Boo a prime target for seeking light-curve anomalies from an edge-on debris disk.

Here we present photometric observations carried out with the CHEOPS space telescope along with spectroscopic measurements with the Hungarian 1m telescope at Piszkéstető Mountain Station and with the SOPHIE spectrograph at the Observatoire de Haute-Provence (OHP). In Section 2 we describe our observations and the data-processing methods we used. In Section 3 we analyze the light curve and present our results. In Section 4 we derive our conclusions and provide a possible explanation of the approximately half-day transient features seen in the residual light curve of DE Boo.

\section{Observations and data processing}

\begin{table}[]
    \centering
    \begin{tabular}{c c c}
    \hline
    \hline
    \noalign{\smallskip}
    Parameter & Value & Reference\\
    \hline
    \noalign{\smallskip}
    $M_A$     & 0.84 M$_\sun$ & \cite{Marshall2014}\\
    $R_A$     & 0.86 R$_\sun$& \cite{Marshall2014}\\
    $L_A$     & 0.498 L$_\sun$& \cite{Marshall2014}\\
    $P_{rot, A}$ & 10.39$\pm$0.03$^\mathrm{d}$ & \cite{Henry1995}\\
    $M_B$     & 0.45 M$_\sun$ & \cite{Kennedy2014}\\
    \noalign{\smallskip}
    \hline
    \hline
    \noalign{\smallskip}
    $R_A$     & 0.870$\pm$0.021 R$_\sun$ & This work\\
    $R_B$   & 0.37$\pm$0.17 R$_\sun$ & This work \\
    $T_{eff,B}$ & 3596$\pm$124 K & This work \\
    $L_B$   & 0.021$_{-0.019}^{+0.029}$ L$_\sun$ & This work \\
    \noalign{\smallskip}
    \hline
    \noalign{\smallskip}
    \end{tabular}
    \caption{Stellar parameters of the K0V primary (A) and the M dwarf companion (B) from the
    literature and this work which were taken as inputs to
    the present analysis. For a detailed description of the estimation of stellar parameters calculated in this present analysis, see Section 3.1.}
    \label{tab:star_param}
\end{table}

\begin{table*}
    \centering
    \begin{tabular}{cccccccc}
    \hline
    \hline
    \noalign{\smallskip}
    Visit & Start Date & End Date & File Key & CHEOPS & Integ. & Num. of     \\
     \# &  &  & & product & time (s) & frames   \\
    \noalign{\smallskip}
    \hline
    \noalign{\smallskip}
    1 & {\small 2021-05-31 17:56:17} & {\small 2021-06-02 18:53:04} & {\small PR100010\_TG001101} & {\it Imagettes} & 3 & 2803 \\
    
    2 & {\small 2021-06-04 20:52:07} & {\small 2021-06-06 21:57:03} & {\small PR100010\_TG001102} & {\it Imagettes} & 3 & 2692 \\
    \noalign{\smallskip}
    \hline
    \noalign{\smallskip}
    \end{tabular}
    \caption{Logs of 2021 CHEOPS observations of DE Boo. The time notation follows the ISO-8601 convention. The File Key supports the fast identification of the observations in the CHEOPS archive.}
    \label{tab:cheops_log}
\end{table*}

During the 2021 visibility window, DE Boo was observed twice with CHEOPS, in May and June. Simultaneous spectroscopic measurements were also gathered from Piszkéstető Mountain Station, Hungary.

\subsection{Photometric observations with CHEOPS}
Observation windows with a length of 30 CHEOPS orbits (1 orbit = 98.77 minutes) were set up for both visits in the CHEOPS Proposal Handling Tool (PHT). The total time of the observation was four days. Observation logs are shown in Table \ref{tab:cheops_log}. As DE Boo is an active star with a brightness of G=5\fm766 in the Gaia G-band \citep{GAIAEDR3}, we used a short exposure time of 3 seconds in order to be able to resolve possible flares and to reduce the effect of saturation. The nominal efficiency of the observations was 71$\%$ and 68$\%$  during the first and second visits to DE Boo. The following ephemeris was used to phase the observations: 
\begin{equation}
    \label{eq:ephem}
    \mathrm{JD}=2459335.084+10.39\times E,
\end{equation}

where 10.39 days was the rotational period of DE Boo.

Photometry was performed using the \textit{imagettes} for each exposure. The imagettes are small images centered on the target with a radius of 30 pixels. The advantage of these images is that because they are smaller in size, they do not have to be co-added on board as in subarrays, and can be downloaded individually. Using the imagettes enabled us to achieve better time resolution, which is crucial in order to adequately deal with flares. Aperture photometry, on the other hand, which is provided by the CHEOPS Data Reduction Pipeline (DRP, \citealt{Hoyer2020}) for subarrays, becomes more difficult in the case of imagettes because of the small sizes. To overcome this issue, we used \texttt{PIPE}\footnote{\url{https://github.com/alphapsa/PIPE}} (PSF imagette photometric extraction; Brandeker et al.\ in prep.; see also descriptions in \citealt{sza21} and \citealt{mor21}), a tool specifically developed for photometric extraction of imagettes using point-spread function (PSF) photometry. This allowed us to derive photometry consistent with the DRP and with comparable S/N, account for instrumental effects such as smearing, and take advantage of the shorter cadence of the imagettes, thus providing help with analyzing flares.

To deal with contamination, frames with recognized issues (e.g.,\ strong background, cosmic rays) were dismissed. Only images unaffected by these events were further used for data analysis.

\subsection{Spectroscopic observations and analysis}
A total of 95 spectra were gathered with the echelle spectrograph ($R = 21\,000$) mounted on the 1\,m RCC telescope at Piszkéstető Mountain Station, Hungary, between 29 May and 2 June 2021, in parallel with the CHEOPS observations, with exposure times of 300s and 600s.  The former yielded an average S/N of 83,  while with the latter an average S/N of 160 was reached at 6400 \AA. Data reduction was carried out with the regular IRAF\footnote{\url{iraf.net}} echelle tasks. Wavelength calibration was carried out using ThAr calibration spectra taken between observations. We also had five observations from the SOPHIE spectrograph mounted on the 1.93m telescope at the Observatoire de Haute-Provence, taken between 31 May and 5 June with an exposure time of 600s and a spectral resolving power of R=75\,000, yielding an average S/N of 255.
%The observation log is presented in Table \ref{tab:pszk_log}.

In order to determine precise astrophysical parameters, spectral synthesis was carried out using SME \citep{piskunov_sme} on both datasets. During the synthesis, MARCS models were used \citep{gustafsson_marcs}. Atomic line parameters were taken from the VALD database \citep{kupka_vald}. Macroturbulence was estimated using the following equation \citep{valenti_macro}:

\begin{equation}
    \label{eq:vmac}
    v_{\mathrm{mac}}=\Bigg(3.98-\frac{T_{\mathrm{eff}}-5770 \mathrm{K}}{650 \mathrm{K}}\Bigg)\,\mathrm{km\,s}^{-1}
.\end{equation}

The whole process is described in \cite{kriskovics2019}. The resulting astrophysical parameters for both the Piszkéstető and OHP datasets are summarized in Table \ref{tab:aphyspar}, and are in relatively good agreement. The slight difference in the $v\sin i$ values could be attributed to the fact that lower resolution results in wider instrumental profiles, which are not always easy to disentangle. We also note that based on the flux ratio of the two components (see Table \ref{tab:star_param}), the contribution of the secondary component to the spectra is negligible.

\subsection{Light-curve pre-processing}

Further corrections were carried out in the photometric data to account for the effects of the systematic errors on roll angle. These errors were corrected following a nonparametric estimation from the data, whereby photometric points were phased according to the roll angle. The data points were smeared in the time domain first, in a boxcar of ten data points in length. After phasing the data in the roll-angle domain, a 2.5 sigma-clipping was applied to omit the outliers, such as possible flares and light-curve transients. The phased light-curve points were then averaged in  roll-angle bins of 3$^\circ$ in width, and a prediction of the roll-angle effect was made for all measured points with a linear interpolation from the binned pattern. 

During the first visit, the star showed flaring activity on two occasions. We plotted these flares in magnification, and find that the noise in the individual data points is compatible with the full amplitude of the flares. This can hide the internal structure of flares if there are multiple flares in succession. To be able to resolve the flare complexes, the light curve was resampled with a cadence of  15 seconds, which meant binning five data points together. This gave the most informative resolution of the flare structures. We also kept this binning in the spot modeling and the analysis of the residuals.

\subsection{Stellar spot model}
\begin{figure*}
    \centering
    \includegraphics[width=2\columnwidth]{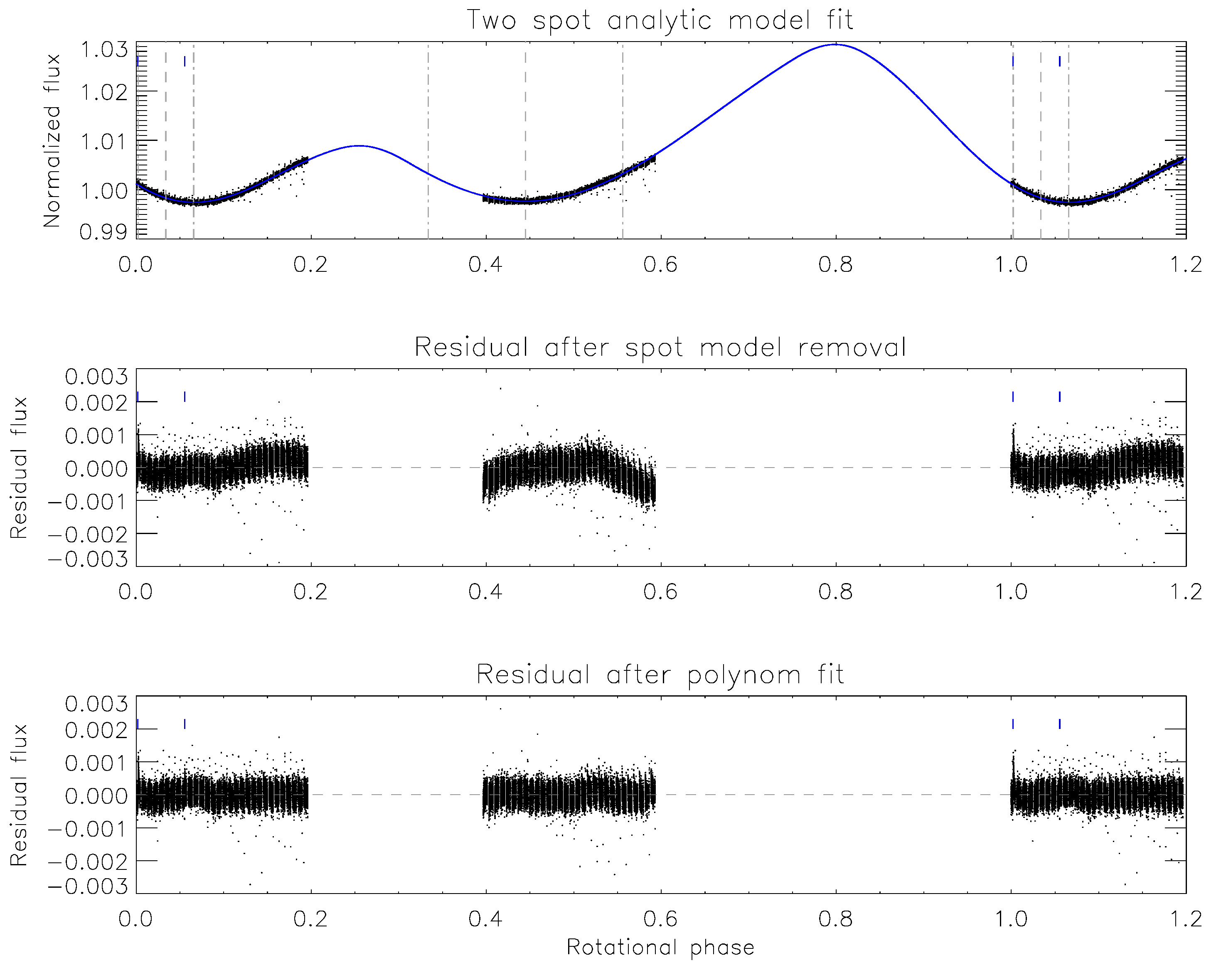}
    \caption{{\Adam Analytical spot model fitted to our CHEOPS data.} Upper panel: CHEOPS light curve (black dots) with the fitted analytic spot model (blue solid line). Dashed lines denote the longitudes of the fitted spots while dash-dotted lines indicate the radii of these spots. Middle panel: Residual after subtraction of the spot model. Lower panel: Residual after  subtraction of the low-order polynomial. Blue ticks denote the positions of the two observed flares. The light curve is phased following Eq. \ref{eq:ephem} with $P_{\mathrm{rot}}=10.39\,\mathrm{d}$}.
    \label{fig:spotmodel}
\end{figure*}

In order to compensate for the rotational modulation, an analytic model of two spots was fitted with \texttt{SpotModel} \citep{ribarik_sml}, following \cite{budding}. The code assumes homogeneous, circular spots with radii, longitudes, and latitudes treated as free parameters, and iteratively changes these parameters until the computed synthetic light curve corresponding to a given spot configuration adequately fits the observed light curve. Assuming $\Delta T \approx 1500\,\mathrm{K}$ (which can be a typical value for cool dwarfs; e.g. \citealt{2003A&ARv..11..153S}) for the spots and using the effective temperature from the spectral synthesis, a spot intensity of 0.25 was used. The rotational period of 10.39 days was kept constant during the spot modeling. The fit is shown in the upper panel of Fig. \ref{fig:spotmodel}, and  the resulting spot parameters are summarized in Table \ref{tab:spots}. As spot latitudes, sizes, and intensities can mutually compensate for each other, latitudes and sizes should be considered nominal. Due to this degeneracy, different spot latitudes, sizes, and temperatures (and therefore spot intensities) can yield virtually the same fit, but as our primary goal is to remove rotational modulation caused by the spots, and because this does not affect the residuals, the fitted model can definitely be used to compensate for the rotational modulation. In the case of high-precision photometry, it is often hard to perfectly fit the light curve with a relatively low number of analytic spots, and therefore it is possible that a slight trend remains (middle panel in Fig. \ref{fig:spotmodel}). As it is clearly seen in Fig. \ref{fig:spotmodel} that the trends are the product of a slight misfit, a further subtraction of a fourth-order polynomial from the residual data was carried out. To ensure no higher harmonics of the stellar rotation were present after the subtraction of the spot model, we studied the periodogram of the data before fitting the polynomial. The resulting periodogram can be seen in in
Fig.~\ref{fig:period} of the Appendix. 

\begin{figure}
    \centering
    \includegraphics[width=\columnwidth]{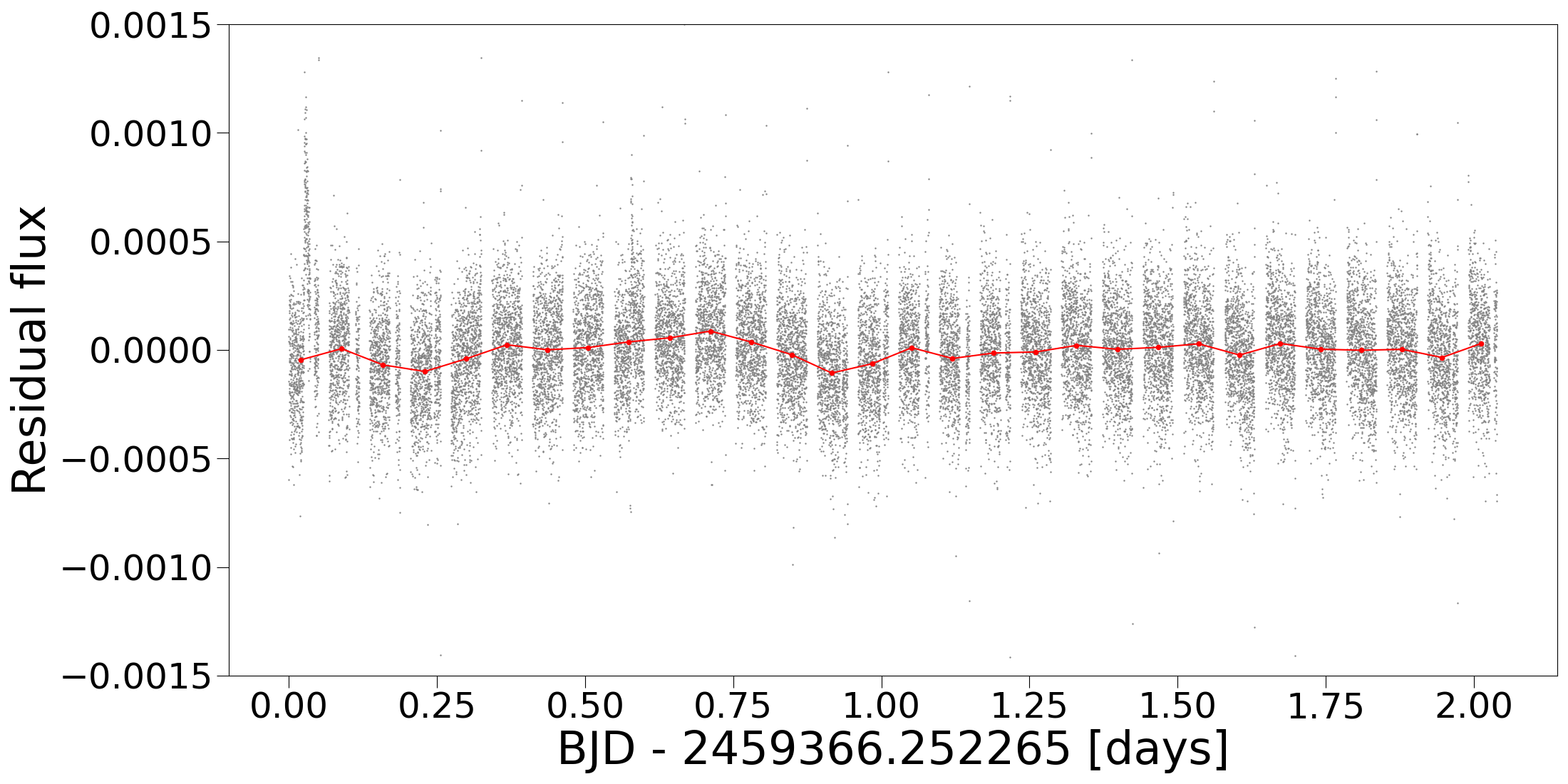}
    \includegraphics[width=\columnwidth]{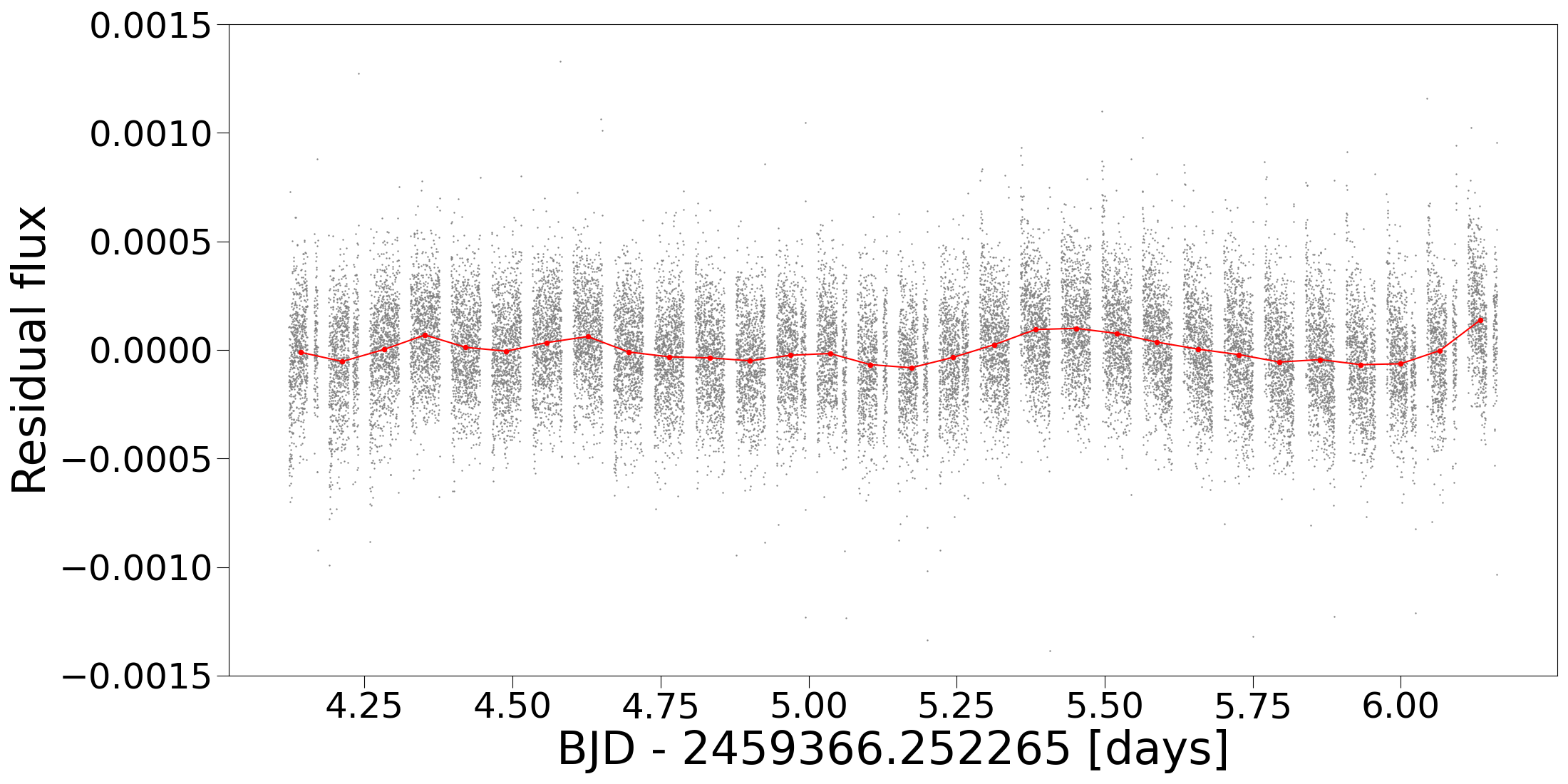}
    \caption{Residual light curves of the first (upper panel) and second (lower panel) visit to DE Boo after the subtraction of a fourth-order polynomial from the spot-model-subtracted data. Red dots show the orbital average flux for each orbit of CHEOPS. Red curves emphasise the change in the brightness of the target.}
    \label{fig:final_res}
\end{figure}

In order to choose the most appropriate model, we performed a Bayesian information criterion (BIC) check over several polynomial models. The fourth-order polynomial model had the lowest BIC value and was used for fitting and further subtraction. 
During this step, we removed the variations on longer timescales ({\Adam $\gtrapprox$1 day} characteristic period), and the order of the applied polynomial has a direct and significant influence on this timescale. In the middle panel of Fig. \ref{fig:spotmodel}, we indeed see variations in this timescale, and these features are very likely not instrumental. CHEOPS is known to be satisfactorily stable in this sense (see e.g., \citet{Delrez2021} or \citet{Morris2021} for the various instrumental effects of CHEOPS). Also, there are no telemeric covectors similar to the observed pattern. However, this timescale does include the light variations caused by the fine structure of the spots on the rotating surface. As we do not have a sufficiently detailed spot model, we cannot make conclusions as to the origin of the long-timescale residuals, or decipher whether they are of a transient nature or are mostly residuals from the rotational signal. Therefore, we decided not to interpret this variation, and removed this signal to reveal the variations on shorter timescales. Here, our aim is not the exact reconstruction of all signals from the disk, but to determine the signal in the data that clearly cannot be of stellar or instrumental origin.
The applied process supported this aim by removing the longer variations seen in the light curve with characteristic periods on the order of days, without affecting the variations on shorter timescales (lower panel of Fig. \ref{fig:spotmodel}), which we discuss in the following sections.
Residual light curves after the subtraction of the polynomial are shown in Fig. \ref{fig:final_res}.

\begin{table}
    \centering
    \begin{tabular}{ccc}
    \hline
    \hline
    \noalign{\smallskip}
       & Piszkéstető  & OHP \\
    \noalign{\smallskip}
    \hline
    \noalign{\smallskip}
$ T_{\mathrm{eff}}$ & $5290\pm93\,\mathrm{K}$ & $5245\pm40\,\mathrm{K}$ \\
$\log g$ &  $4.47\pm0.15$ & $4.5 \pm 0.05$ \\
$[\mathrm{Fe}/\mathrm{H}]$ & $0.09\pm0.1$ & $0.14 \pm 0.07 $ \\
$v_{\mathrm{mic}}$ & $1.3\pm0.9\,\mathrm{kms}^{-1}$ & $1.0\pm0.1\,\mathrm{kms}^{-1}$ \\
$v_{\mathrm{mac}}$ & $4.7\,\mathrm{kms}^{-1} $ & $4.7\,\mathrm{kms}^{-1} $ \\
$v\sin i$ & $7\pm1 \, \mathrm{kms}^{-1}$ & $4.5 \pm 0.5 \, \mathrm{kms}^{-1}$ \\
    \noalign{\smallskip}    
    \hline
    \noalign{\smallskip}
    \end{tabular}
      \caption{Astrophysical parameters from spectral synthesis for the Piszkéstető and OHP datasets.}
    \label{tab:aphyspar}
\end{table}

\begin{table}
    \centering
    \begin{tabular}{ccc}
    \hline
    \hline
    \noalign{\smallskip}
       & Spot 1 & Spot 2 \\
    \noalign{\smallskip}
    \hline
    \noalign{\smallskip}
    $l$ [deg] & $12.17 \pm 0.05$ & $160.34 \pm 0.11$  \\
    $b$ [deg] & $-10.16 \pm 5.30$ & $83.60 \pm 0.05$ \\
    $r$ [deg] & $11.42 \pm 0.21$ & $40.12 \pm 0.03$\\ 
    \noalign{\smallskip}
    \noalign{\smallskip}
  \hline
  \noalign{\smallskip}
    \end{tabular}
    \caption{Spot parameters from the analytical spot model. $l$, $b,$ and $r$ denote spot longitudes, latitudes, and radii, respectively.}
    \label{tab:spots}
\end{table}

\subsection{Estimating flare energies}

\begin{figure}
    \centering
    \includegraphics[viewport=10 122 437 335,width=\columnwidth]{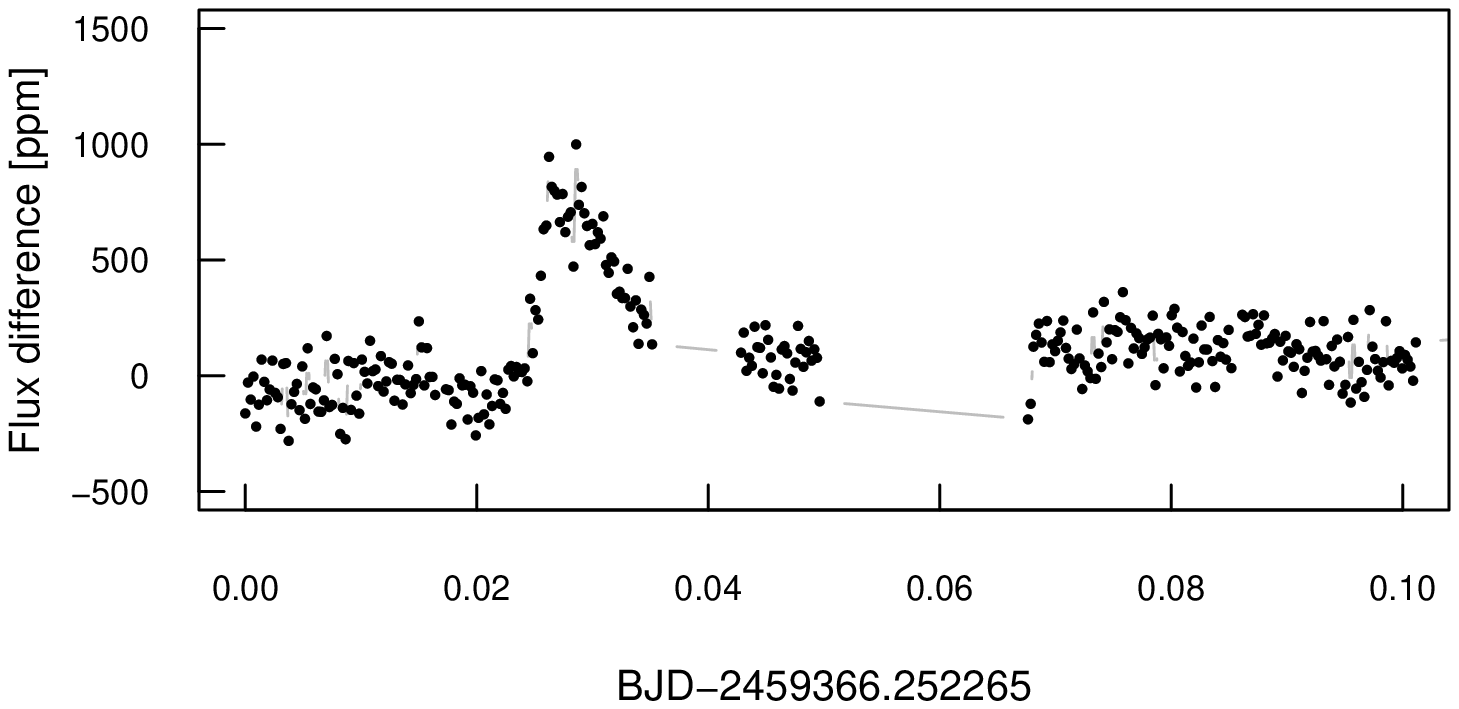}
    \includegraphics[viewport=10 122 437 335,width=\columnwidth]{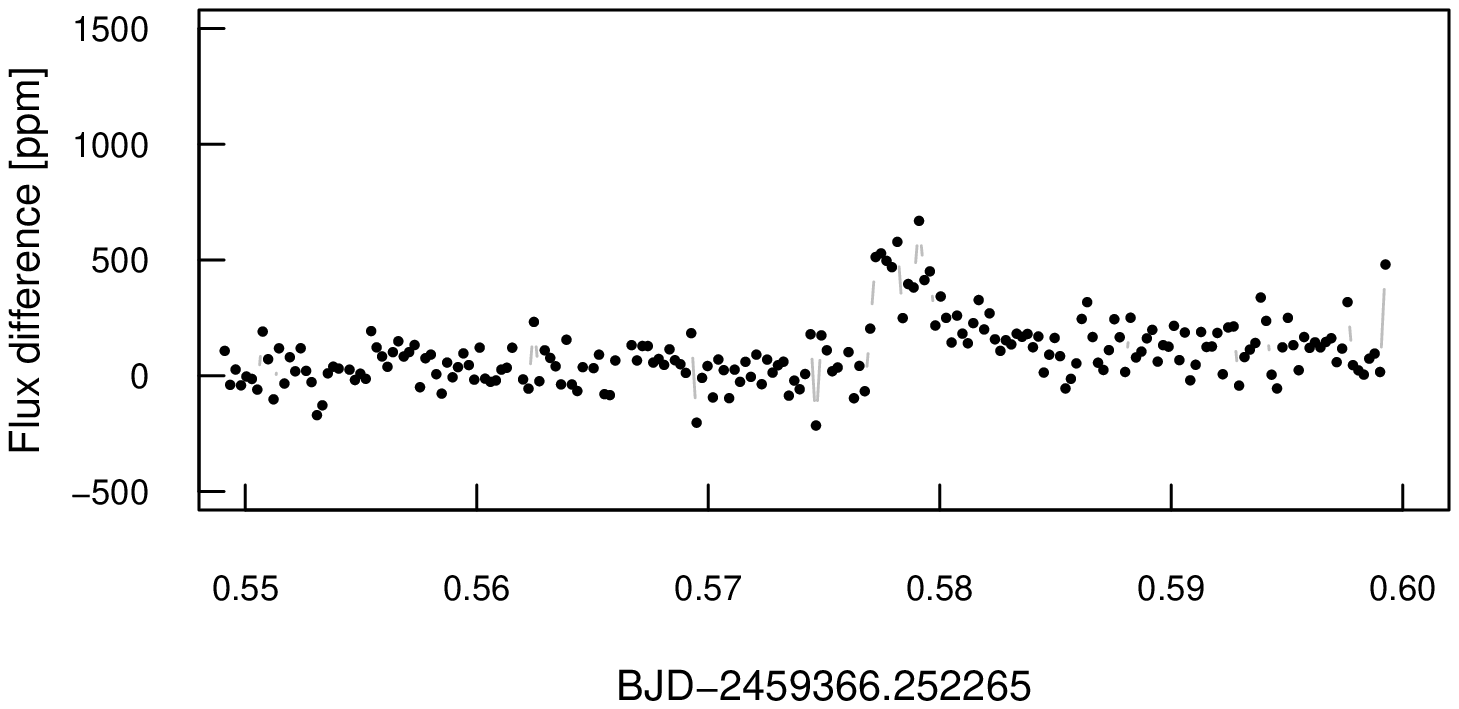}
    \caption{Two flares that occurred during the first visit to DE Boo, after re-sampling the imagette data to a cadence of 15 s.}
    \label{fig:flares}
\end{figure}

During the first visit to DE Boo, two flares were detected by CHEOPS. This is the first time that  flares have been photometrically observed and reported for DE Boo. This unprecedented detection justified a detailed analysis of these phenomena. In order to do so, we needed imagettes of sufficiently short cadence  to resolve the finer structures of the flares. However, the resampling of these data for 15 seconds was also necessary to increase the S/N. The flares in the re-sampled imagette data are shown in Fig. \ref{fig:flares}. 

In order to obtain the fraction of DE Boo's quiescent luminosity in the CHEOPS band, a BT-NextGen model spectrum for a K0V-type star with T$_{\mathrm{eff}}=$5300~K and logg$=4.5$ was convolved with the transmission curve of CHEOPS \citep{Deline2020}. Flare energies were then calculated by integrating the total flux for the duration of the flaring events, and multiplying them with the bolometric luminosity of DE Boo and the fraction of the luminosity in the CHEOPS band. Though we assumed that the flares originate from the primary, in reality they could arise from the M dwarf companion as well. However, as both the quiescent luminosity and the stellar spectrum are dominated by the K0V primary component, our approach is expected to provide a reasonable estimation of the flare energies. We note that the order of magnitude of the energies would not be different even if they originated from the companion.

\section{Results}

\subsection{Improved radius of DE Boo}
We improved existing measurements of the radius of DE Boo using a Markov-Chain Monte Carlo infrared flux method (MCMC IRFM; \citealt{Blackwell1977,Schanche2020}. This approach uses known relations between the stellar angular diameter, effective temperature, and apparent bolometric flux. For DE Boo, we computed the bolometric flux by constructing a spectral energy distribution (SED) of two stellar components: one of the primary using the spectral parameters derived above with the \textsc{atlas} catalogs \citep{Castelli2003} and a second of the known M-dwarf companion with parameters taken from the \textsc{phoenix} catalogs \citep{Allard2011}. Empirical relations from \citet{Baraffe2015} were used to estimate the effective temperature and the radius of the companion. This was necessary in order to derive an accurate stellar radius for DE Boo using the derived parameters and the offset-corrected Gaia EDR3 parallax \citep{Lindegren2021}. We fitted the combined models to observed data taken from the most recent data releases for the following bandpasses: Gaia G, G$_{\rm BP}$, and G$_{\rm RP}$, 2MASS J, H, and K, and {\it WISE} W1 and W2 \citep{Skrutskie2006,Wright2010,GaiaCollaboration2021}. We used this fit to derive $R_{\star}=0.870\pm0.021\, R_{\odot}$.

\subsection{Properties of flare activity} 
We estimated the energies of the two flares observed during the first visit. The first, more energetic flare, which lasted for 20 minutes, occurred at the beginning of the observation period and had an estimated energy of 3.35$\times$10$^{32}$ erg. In contrast, the second flare had a much shorter duration of 5 minutes and an energy of about 7.43$\times$10$^{31}$ erg and was only identifiable on the imagettes. The detection of two flares with such energies under the total observation time of four days is not out of the ordinary on a fairly active late G or early K dwarf (e.g. \citealt[and references therein]{leitzinger2020}). The energy levels are relatively typical as well \citep{vida2022flare}. Assuming that these flares originated from the companion, their energies would still be well within the boundaries of what is expected in the case of an M dwarf star. However, we note that two flares during a four-day period is not enough to make any  quantitative conclusions about the flare activity. Derived parameters of the two flares along with their dates of occurrence are shown in Table \ref{tab:flares}.

\begin{table}
    \centering
    \resizebox{\columnwidth}{!}{\begin{tabular}{c c c c}
    \hline
    \hline
    \noalign{\smallskip}
         \makecell{Start date\\ JD-2400000} & \makecell{End date \\ JD-2400000} & \makecell{Duration \\ $[min]$} & \makecell{Estimated energy \\ $[erg]$}\\
    \noalign{\smallskip}
    \hline
    \noalign{\smallskip}
         59366.273541 & 59366.287429 & 20 & 3.35$\times$10$^{32}$ \\
         59366.830820 & 59366.834061 & 4.7 & 7.43$\times$10$^{31}$ \\
    \noalign{\smallskip}
    \hline
    \noalign{\smallskip}
    \end{tabular}}
    \caption{Estimated energies and durations of the two flares during the first visit to DE Boo.}
    \label{tab:flares}
\end{table}

\subsection{Possible photometric transients in the light-curve residuals}

After masking out the flares, subtracting a spot model, and removing slowly varying systematic errors due to possible structures on the stellar surface that were not accounted for in our simple two-spot model, we noticed the presence of moderately slowly varying light-curve residuals on a timescale of 0.3--0.8 days. These residuals have a shorter timescale than what could be originated by the rotation of spots (and the timescale of spot emergence or decay is much longer than this; see e.g. \citealt{namekata}), while the timescale is too long to be due to flares or other activity transients. Other typical light curve changes, such as ellipticity in closer binaries, or other tracers of magnetic activity typical of RS\,CVn systems are all related to rotation, and in our case would cause changes on longer timescales \citep[see e.g.][and references therein]{berd2005}. Therefore, we interpret these residuals as transients from the edge-on disk in the following.  

Let us assume a cylindrically symmetric disk consisting of a light scattering material that follows some $\sigma(r,\phi)$ distribution. Here, $\sigma$ is the scattering cross section of the disk locally, that is, the fraction of starlight that is occulted or scattered by the dust particles in the elementary volume; $r$ is the radius vector, and $\phi$ is the angular coordinate. If the dust follows a circular orbit, the $\omega$ angular velocity can be expressed as $\omega=\sqrt{2GM} r^{-3/2}$, according to Kepler's Third Law. Hence, during a $\rm{d}t$ time, we can write
\begin{equation}
{\rm d}\phi = 
\omega {\rm d}t =
\sqrt{2GM} \,   r^{-\frac{3}{2}}\, {\rm d}t.
\end{equation}

The variation of the disk scattering along the line of sight is then
\begin{equation}
\frac{{\rm d}\sigma}{{\rm d}t} =
\frac{{\rm d}\sigma}{{\rm d}r}
\frac{{\rm d}r}{{\rm d}t} +
\frac{{\rm d}\sigma}{{\rm d}\phi}
\frac{{\rm d}\phi}{{\rm d}t} =
\frac{{\rm d}\sigma}{{\rm d}\phi} \sqrt{2GM} \, r^{-\frac{3}{2}}\,
,\end{equation}
because $\frac{{\rm d}r}{{\rm d}t}=0$. 
The variation of the total scattering cross section along the line of sight $\Sigma=\int_{r1}^{r2} \sigma {\rm d}r$ can then be expressed as
\begin{equation}
\frac{{\rm d}\Sigma}{{\rm d}t} =
\int_{r1}^{r2}
\frac{{\rm d}\sigma}{{\rm d}\phi} \sqrt{2GM} \, r^{-\frac{3}{2}}\,
{\rm d}r.
\end{equation}

This formula represents the amount of material along a distinct
line of sight. The light-curve transients are related to the transits
of the individual grains along the stellar disk. If a material at $r$ radius with $\rho$ density scatters out a given fraction of the light  $S$, the contribution to the light curve can be written as 
\begin{equation}
S(r,t)=\left[ \rho(r,\tau) \ast Lc(r,\tau) \right] (t),
\end{equation}
where $Lc$ is the transit light curve of a compact transiter
at $r$ radius, $\ast$ denotes convolution, and $\tau$ is a running time-like variable, which was introduced because the transit light curve is a time-dependent variable itself. This expression shows that the shape of the transit light curve can be derived as the shapes of all individual overdensities at their $r$ radius and $\tau$ transit times combined appropriately and evaluated at t. If the configuration is very thin, and the scattered light is on the order of one part per thousand at most, the combination can be simplified to a summation. In this case, the combined effect of all contributions to the light curve shape will be written as
\begin{equation}
S(t)=\int_{R1}^{R2} \, 
\left[ \rho(r,\tau) \ast Lc(r,\tau) \right] (t) \, dr.
\end{equation}

Here, the convolution shows that the frequency spectrum of the
transit light curve will be the $F(\sigma)$ {\Adam frequency spectrum} belonging to the material distribution multiplied by the $F(T)$ {\Adam frequency spectra} of the transit light curves at different radii. If the cloud of grains is compact (the transiting structure is much smaller than the stellar disk) and $F(\sigma)$ includes components at very large frequencies, the boundary frequency in F(signal) will be the highest frequency in $F(T)$. If the transiting feature is compact, for example a
small planet, the duration of the transient will be close to the transit duration. If the transiting feature is extended, for example a loose cloud of large asteroids or a giant comet, where the ingress and egress duration are compatible with (or even exceed) the transit duration, the experienced timescale of the transients will be significantly longer than the transit duration calculated at the orbital semi-major axis of the transiting feature.

If the transiting cloud is extended far beyond the stellar radius, $(\sigma)$ has only low-frequency components, and the observed light-curve anomaly will be very slow, on the order of $\Delta \phi/\sqrt{(2GM)}r^{-\frac{3}{2}}$.
In this case, the smoothness of the distribution of the material and the orbital radius of the disk features become degenerate. Therefore, giving a lower limit for the orbital radius of the excess material, we also have to give an upper limit for its angular extension. Similar descriptions of these processes can be found in \citet{Kennedy2017}.

In summary, the transit of the grains that generate a light-curve anomaly can be at most as long as the anomaly itself. Applying Kepler's Third Law, this gives us an estimate, or more precisely, an upper limit of the radius where the claimed transiting features can be found, which we can write as
\begin{equation}
a_{max} = \frac{GM_\star D^2}{4 (1-b^2) R_{\star}^2},
\label{distance}
\end{equation}
where G is the gravitational constant, M$_\star$ and R$_\star$ are the mass and radius of DE Boo, respectively, $D$ is the duration of the light-curve anomaly itself, measured in units of time, and $b$ is the impact parameter, the projected distance of the transit chord from the center point of the star, measured in units 
of $R_{\star}$. If the shape of the transient is compatible with the transit of a compact, planet-like object, $D$ is related to the relative duration $W$, as $D=W\times P_{orb}$, where $W = R_{\star}/a\sqrt{(1+k)^2-b^2}/\pi$. If the light variation is more fuzzy and we cannot define a major transit bin, $D$ is to be understood as simply the time from the beginning to the end of the negative transit anomalies. 

The residuals of the DE Boo observations are plotted in Fig.~\ref{fig:final_res}. The residuals of the two visits differ from each other. The first visit is compatible with a stable residual, with the dips suggesting one or two transients passing in front of the star. The most stable part of the residual is observed between 2459367.3 and 2459368.3. The second CHEOPS visit is merely composed of varying residuals and suggests density variations around the star rather than one transiting object. 
The most significant deviations from a constant residual
are observed between 2459367.0 and 2459367.3, 2459370.8 and 2459371.6, and 2459371.75 and 2459372.35.
Therefore, the experienced timescale of the suspected transients in the DE Boo light curve is 0.3--0.8 days.
 
Using the stellar parameters given in Table \ref{tab:star_param} and inserting 0.3 and 0.8 days for the transit duration into Eq. \ref{distance} and assuming $b=0$, we get 0.35 and 2.47 AU for the maximum orbital distance of the transiting feature. This places the transiting object much closer to the star than the debris disk, to indicatively 1~AU from DE Boo. 
The suspected structure is consistent with the
criteria of dynamical stability according to \cite{1999AJ....117..621H}. This criterion tells us that
circumbinary structures farther from the system center than the $a_c$ critical radius are stable in the long term; whereas values of $a_c$ were found by means of numerical simulations and as a function of the $\mu$ mass ratio and the eccentricity of the binary. Table 7 in \cite{1999AJ....117..621H} shows us that near $\mu=$0.4--0.5, $e=0.5$ and $a_c$=3.6--3.7 if measured in the unit of the primary-to-secondary separation. As this separation is 0.19~AU, debris structures behind the critical radius of $a_c=0.703$~AU can be dynamically stable. Therefore, a transient at $~1$ AU from the barycenter is dynamically plausible. We also note that \cite{2007ApJ...658.1289T} review several other dynamical criteria, but these authors find 
the \cite{1999AJ....117..621H} criterion to be the strictest of these, and therefore we applied the \cite{1999AJ....117..621H} criterion here.

\section{Discussion and conclusions}

We carried out photometric observations of DE Boo  using the CHEOPS space telescope with a total duration of four days. In order to properly remove the effect of stellar variation, a stellar spot model and a fourth-order polynomial were fitted and subtracted from the light curves, respectively. We observed two flares during the observation period. This was the first time that flaring activity of DE Boo was reported. We found the estimated energies of the flares to be consistent with the predicted levels of activity of a magnetically active K dwarf star. We conclude that further measurements are required to obtain quantitative results about the stellar activity of DE Boo. 

{\Adam Local Thermodynamic Equilibrium (LTE)} synthetic spectra were fitted to the observations taken with the echelle spectrograph mounted on the  1m RCC telescope at Piszkéstető, Hungary, and with SOPHIE installed on the 1.93m telescope at the Observatoire de Haute-Provence, France, in order to refine the astrophysical parameters of the primary. The two fits yielded {\Adam the same} results within their respective error bars.

Nonperiodic transiting features with a duration of 0.3--0.8 days were identified in the residual light curves of DE Boo. Based on their transit duration, the semi-major axes of these features were estimated. Our calculations show that these features had a maximum orbital distance of 2.47 AU.
This is a much smaller value than the indicated inner radius of the debris disk that is seen in the infrared wavelengths. As the duration of our observations was less than the expected transit duration of the material passing in the debris disk at 70~AU according to Eq.~\ref{distance} (around 4.5~days), it is not possible to detect their effects on the investigated light curves. Space telescopes with longer observational baselines (such as TESS) are better suited to this task. Moreover, we note that any effect with the duration of a few days was removed during the polynomial fitting. However, the goal of this study was not simply to identify features of an already known debris disk. As it is not out of the ordinary that an inner disk is found in a system with outer debris disks, it cannot be excluded that,  for DE Boo, transiting material is observed much closer to the star than the most massive parts of the debris disk. In the case of $\beta$ Pic, we see a similar geometry: although the visible debris disk
extends to 100 (secondary disk) and $~$200 (primary disk) AU, the observed transients last for less than a day, indicating that the presumed exocomets orbit much closer than the debris disks. Several systems with warm debris disks have also been known to exhibit quasi-periodic dippers with a timescale of 1--4 days; these were found to be due to transits of overdensities in the disk, of surprisingly large amplitude on some occasions \citep{2020ApJS..251...18T}. These systems are more similar to DE Boo in terms of the origin of the light variations. In these systems, the overdensities also appear in a structure that is close to the star (as can be concluded from their quasi-period), while the major warm debris disk lies much farther out.

A possible explanation {\Adam for the observed light variations} is the formation of an inner disk region, indicating ongoing dust replenishment \citet{Wahhaj2003}, which can have an origin in the planetesimal belts \citet{Okamoto2004}. Warm debris disks around evolved stars cannot be primary remnants of planet formation, but formed quite recently \citep{Wyatt2007a}, most likely in a late heavy bombardment-like scenario, or from a very massive reservoir of planetesimals that survived planet formation (such as the Oort Cloud in the Solar System). In the presence of warm dust, excess radiation 
is expected in {\Adam mid-infrared}. However, no excess was detected at 24~$\mu$m by the Multiband Imaging Photometer for Spitzer (MIPS) for DE Boo \citep{Gaspar2013}. This may be because the detection limit of MIPS for debris disks at a distance of a few AU around late-type stars is above $f\sim10^{-5}$. However, variations in the light curve of DE Boo imply a lower fractional luminosity, which is therefore below what may be detectable with MIPS. This suggests that the presence of a faint warm debris disk cannot be excluded.

The current observations covered 2$\times$2 days of
DE Boo, and show that DE Boo is an important example of a system with an edge-on debris disk that we are looking at through the debris disk itself. Further observations with TESS in 2022, covering 28 days, will be an ideal opportunity to confirm our interpretations.

\begin{acknowledgements}

CHEOPS is an ESA mission in partnership with Switzerland with important contributions to the payload and the ground segment from Austria, Belgium, France, Germany, Hungary, Italy, Portugal, Spain, Sweden, and the United Kingdom. The CHEOPS Consortium would like to gratefully acknowledge the support received by all the agencies, offices, universities, and industries involved. Their flexibility and willingness to explore new approaches were essential to the success of this mission. 
We acknowledge the support of the PRODEX Experiment Agreement No. 4000137122 between the ELTE Eötvös Loránd University and the European Space Agency (ESA-D/SCI-LE-2021-0025). 
GyMSz acknowledges the support of the Hungarian National Research, Development and Innovation Office (NKFIH) grant K-125015, a a PRODEX Experiment Agreement No. 4000137122, the Lend\"ulet LP2018-7/2021 grant of the Hungarian Academy of Science and the support of the city of Szombathely. 
LK is supported by the Hungarian National Research, Development and Innovation Office grants PD-134784 and K-131508. LK is a Bolyai János Research Fellow. 
ABr was supported by the SNSA. 
FK acknowledges funding from the Centre National d'Etudes Spatiales, from the French National Research Agency (ANR) under contract number ANR-18-CE31-0019 (SPlaSH), and from the Université Paris Sciences et Lettres under the DIM-ACAV program Origines et conditions d'apparition de la vie. 
DG gratefully acknowledges financial support from the CRT foundation under Grant No. 2018.2323 ``Gaseous or rocky? Unveiling the nature of small worlds''. 
LMS gratefully acknowledges financial support from the CRT foundation under Grant No. 2018.2323 ‘Gaseous or rocky? Unveiling the nature of small worlds’. 
ACC and TW acknowledge support from STFC consolidated grant numbers ST/R000824/1 and ST/V000861/1, and UKSA grant number ST/R003203/1. 
S.G.S. acknowledge support from FCT through FCT contract nr. CEECIND/00826/2018 and POPH/FSE (EC). 
YA and MJH acknowledge the support of the Swiss National Fund under grant 200020\_172746. 
We acknowledge support from the Spanish Ministry of Science and Innovation and the European Regional Development Fund through grants ESP2016-80435-C2-1-R, ESP2016-80435-C2-2-R, PGC2018-098153-B-C33, PGC2018-098153-B-C31, ESP2017-87676-C5-1-R, MDM-2017-0737 Unidad de Excelencia Maria de Maeztu-Centro de Astrobiologí­a (INTA-CSIC), as well as the support of the Generalitat de Catalunya/CERCA programme. The MOC activities have been supported by the ESA contract No. 4000124370. 
S.C.C.B. acknowledges support from FCT through FCT contracts nr. IF/01312/2014/CP1215/CT0004. 
XB, SC, DG, MF and JL acknowledge their role as ESA-appointed CHEOPS science team members. 
ACC acknowledges support from STFC consolidated grant numbers ST/R000824/1 and ST/V000861/1, and UKSA grant number ST/R003203/1. 
This project was supported by the CNES. 
The Belgian participation to CHEOPS has been supported by the Belgian Federal Science Policy Office (BELSPO) in the framework of the PRODEX Program, and by the University of Liège through an ARC grant for Concerted Research Actions financed by the Wallonia-Brussels Federation. 
L.D. is an F.R.S.-FNRS Postdoctoral Researcher. 
This work was supported by FCT - Fundação para a Ciência e a Tecnologia through national funds and by FEDER through COMPETE2020 - Programa Operacional Competitividade e Internacionalizacão by these grants: UID/FIS/04434/2019, UIDB/04434/2020, UIDP/04434/2020, PTDC/FIS-AST/32113/2017 \& POCI-01-0145-FEDER- 032113, PTDC/FIS-AST/28953/2017 \& POCI-01-0145-FEDER-028953, PTDC/FIS-AST/28987/2017 \& POCI-01-0145-FEDER-028987, O.D.S.D. is supported in the form of work contract (DL 57/2016/CP1364/CT0004) funded by national funds through FCT. 
B.-O.D. acknowledges support from the Swiss National Science Foundation (PP00P2-190080). 
This project has received funding from the European Research Council (ERC) under the European Union’s Horizon 2020 research and innovation programme (project {\sc Four Aces}. 
grant agreement No 724427). It has also been carried out in the frame of the National Centre for Competence in Research PlanetS supported by the Swiss National Science Foundation (SNSF). DE and AD acknowledge financial support from the Swiss National Science Foundation for project 200021\_200726. 
MF and CMP gratefully acknowledge the support of the Swedish National Space Agency (DNR 65/19, 174/18). 
M.G. is an F.R.S.-FNRS Senior Research Associate. 
SH gratefully acknowledges CNES funding through the grant 837319. 
KGI is the ESA CHEOPS Project Scientist and is responsible for the ESA CHEOPS Guest Observers Programme. She does not participate in, or contribute to, the definition of the Guaranteed Time Programme of the CHEOPS mission through which observations described in this paper have been taken, nor to any aspect of target selection for the programme. 
This work was granted access to the HPC resources of MesoPSL financed by the Region Ile de France and the project Equip@Meso (reference ANR-10-EQPX-29-01) of the programme Investissements d'Avenir supervised by the Agence Nationale pour la Recherche. 
ML acknowledges support of the Swiss National Science Foundation under grant number PCEFP2\_194576. 
PM acknowledges support from STFC research grant number ST/M001040/1. 
LBo, GBr, VNa, IPa, GPi, RRa, GSc, VSi, and TZi acknowledge support from CHEOPS ASI-INAF agreement n. 2019-29-HH.0. 
This work was also partially supported by a grant from the Simons Foundation (PI Queloz, grant number 327127). 
IRI acknowledges support from the Spanish Ministry of Science and Innovation and the European Regional Development Fund through grant PGC2018-098153-B- C33, as well as the support of the Generalitat de Catalunya/CERCA programme. 
V.V.G. is an F.R.S-FNRS Research Associate. 
NAW acknowledges UKSA grant ST/R004838/1.

\end{acknowledgements}

\bibliographystyle{aa} % style aa.bst
\bibliography{DEBoo_ref} % your references Yourfile.bib

\begin{appendix}
\section{Periodogram of spot-model-subtracted data}

\begin{figure}[h]
    \centering
    \includegraphics[width=\columnwidth]{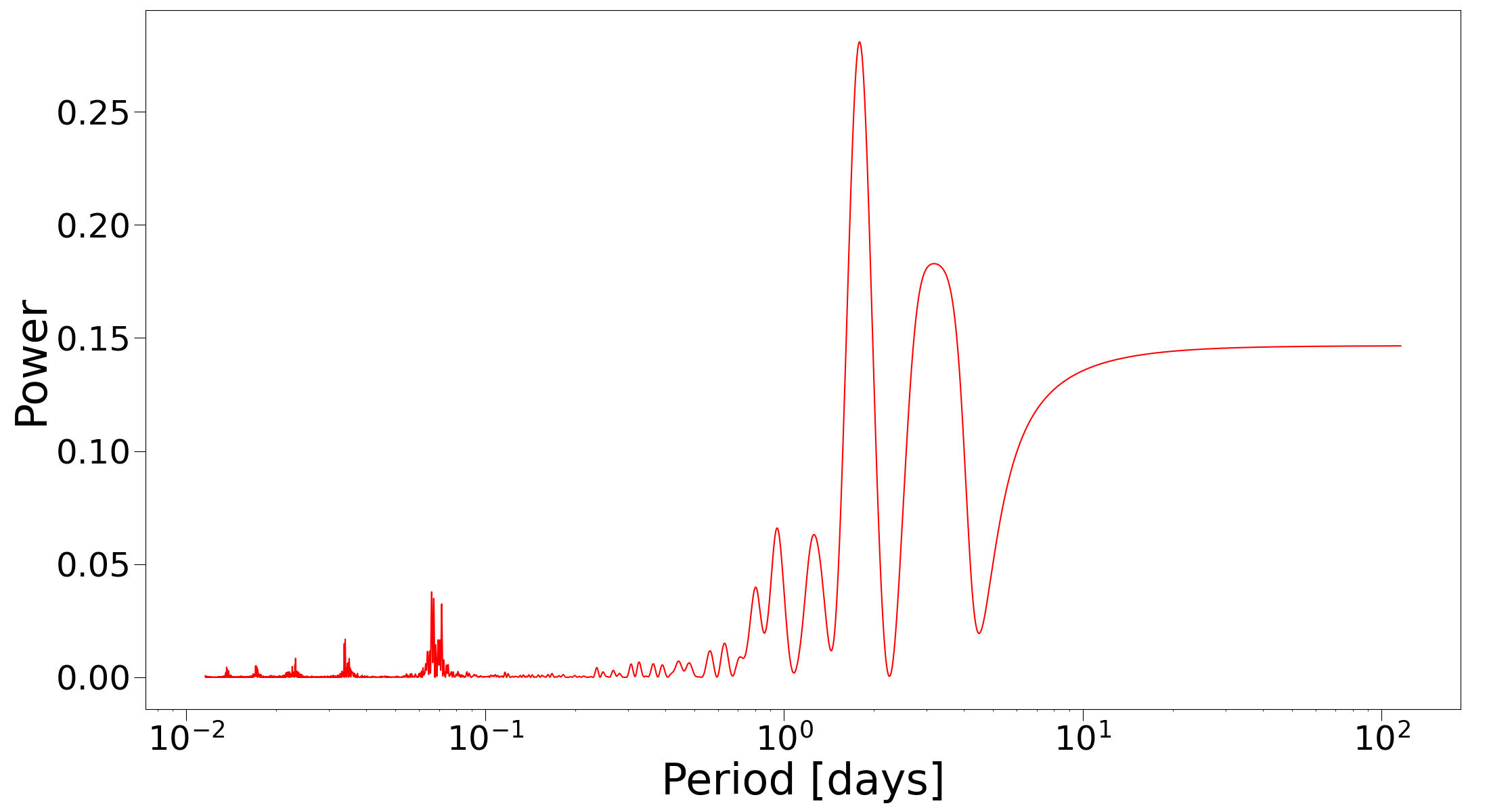}
    \caption{Lomb-Scargle periodogram of the spot-model-subtracted data. The peak with the maximum power belongs to the period of 1.79~days. We note that the effect of the 98.77 minute rotational period of CHEOPS also appears as a peak around 100~minutes.}
    \label{fig:period}
\end{figure}

We applied the Lomb-Scargle periodogram to the spot-model-subtracted data from both visits before applying the fourth-order polynomial to check for remaining periods. Figure~\ref{fig:period} shows the resulting periodogram, where the strongest peak was seen at 1.79 days. Another peak was present at around 100~minutes due to the orbital period of CHEOPS. We concluded that no higher harmonics of the stellar rotation of 10.39 days were present in the spot-model-subtracted data.

\end{appendix}

\end{document}